# A review of conventional explanations of anomalous observations during solar eclipses

Chris P. Duif

Faculty of Applied Sciences, Department of Radiation, Radionuclides & Reactors
Delft University of Technology,
Mekelweg 15, 2629 JB, Delft, The Netherlands
The Netherlands
eMail: c.p.duif@iri.tudelft.nl / chrisd@space-time.info

**Abstract**

Conventional explanations for observations of anomalous behaviour of mechanical systems during solar eclipses are critically reviewed. These observations include the work of Allais with paraconical pendula, those of Saxl and Allen with a torsion pendulum and measurements with gravimeters. Attempts of replications of these experiments and recent gravimeter results are discussed and unpublished data by Latham and by Saxl et al. is presented. Some of the data are summarized and re-analyzed. Especially, attention is paid to observations of tilt of the vertical, which seems to play an important role in this matter and recommendations for future research are given. It is concluded that all the proposed conventional explanations either qualitatively or quantitatively fail to explain the observations.

PACS numbers: 04.80Cc, 4.80.Nn, 95.10.Gi, 91.10.Pp, 04.80.Cc, 94.10.Dy, 92.60.Fm, 92.60.Dj, 91.30.Dk, 91.10.Kg

## 1. Introduction

In recent years there has been a renewed interest in reports about anomalies during solar eclipses. Realizing that our understanding of gravity at galactic scales may be insufficient (giving rise to theories like MOND [Mil83, SanM02]), the observation of an anomalous acceleration on spacecrafts in the solar system (the Pioneer Anomaly [AndL02]), anomalous velocity increases of spacecrafts during Earth flybys [AntG98, AndW00, NieT04] and even discussions about whether we understand gravity at laboratory scale [MbeL02, Mel99, GerG02] may have contributed to this renewed interest. Also the (unfinished) project by NASA [NASA99] in 1999 played a role as well as several publications which hypothesized conventional causes for the eclipse phenomena [UnnM01, FlaY03, RuyS03, YanW02].

Since the 1950's there have been several reports on anomalous behaviour of various kinds of mechanical oscillators during solar eclipses. Replications of these experiments by others did not always confirm the existence of these effects. Concerning these replications, it can be remarked, however, that most of these were modified versions of the original experimental setups and that the conditions of various solar eclipses (elevation of the Sun during the eclipse, geographical latitude, speed of the shadow, etc.) differ remarkably from one to another. Little attention has been paid to these differences because often only simple classical mechanical mechanisms (*e.g.,* Majorana shielding of gravity [Eck90, Edw02, Fla96, SliC65]) were considered.

The first publications, by Allais, were about measurements with so-called paraconical pendula [All57, All59, All97]. These reports are still considered to be among the most reliable, not in the least because the results of measurements during two solar eclipses were qualitatively and quantitatively similar (see Figs. 1 and 2). Later, also reports about observations with torsion pendula [SaxA71] and with Foucault pendulum [MihM03] were added to this. Several replications of these experiments yielded negative results [Kuu91, Kuu92, JunJ91]. A re-analysis of the observations of Saxl and Allen during the 1970 solar eclipse in the USA is depicted in Fig. 3.

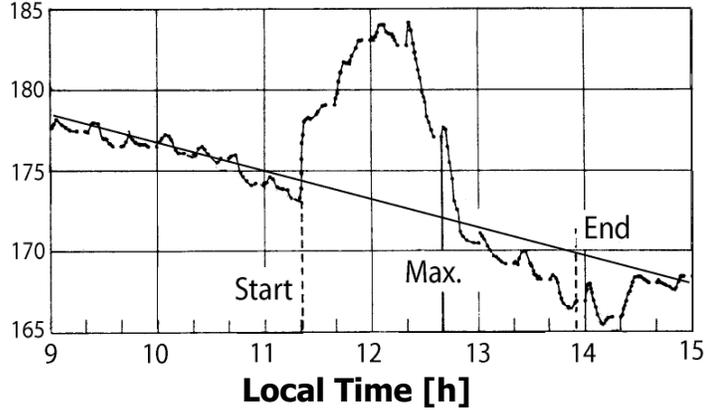

**Fig. 1.** Measurements during the June 30, 1954 solar eclipse with a paraconical pendulum by Allais [All59a, All97]. Vertical axis is the azimuth angle of the pendulum trajectory (in grades, 100 grades=90º). Also a linear trend line is shown. Each point is the accumulated effect of 14 min of measurement, pendulum release is every 20 min.

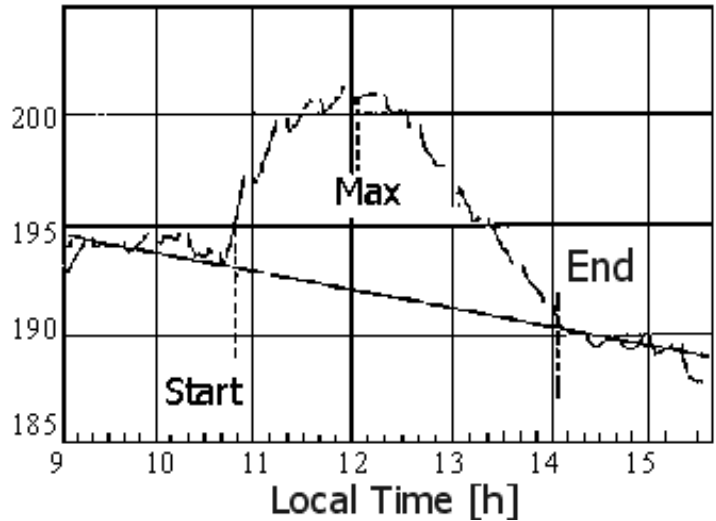

**Fig. 2.** Measurements during the 2 October 1959 solar eclipse with a paraconical pendulum by Allais [All97]. Vertical axis is the azimuth angle of the pendulum trajectory (grades).

Latham noticed that the anomalous deviations during the eclipse of the paraconical pendulum of Allais had the same time behaviour as that of the torsion pendulum of Saxl and Allen (change in azimuth angle and change in oscillation period, respectively) [Lat80]. The quantity measured by Allais is actually an integral value, the action building up during the 14 min of measurements. When one fits the Allais data with a smooth curve and takes its derivative, a time behaviour like depicted in Fig. 4 becomes visible. These time-patterns may be important in finding the cause(s) of the effects.

Recently, some positive reports about observations with gravimeters during solar eclipses by a Chinese team [WanY00, YanW02, TanW03] attracted attention.

Decreases in the gravitational acceleration up to ~7 μgal, just before 1st contact and around the 4th contact were measured (1 μgal = $10^{-8}$ $ms^{-2}$). These measurements were performed after a positive result by Mishra and Rao during an eclipse in 1995 in India [MisR97], see Fig. 5. Measurements with gravimeters by others, however, were all negative [DucS99, Meu00, Man01, Tom55]. These earlier negative results with gravimeters led to speculations that the effects were only observable with dynamical measurement devices like torsion pendula [SaxA71].

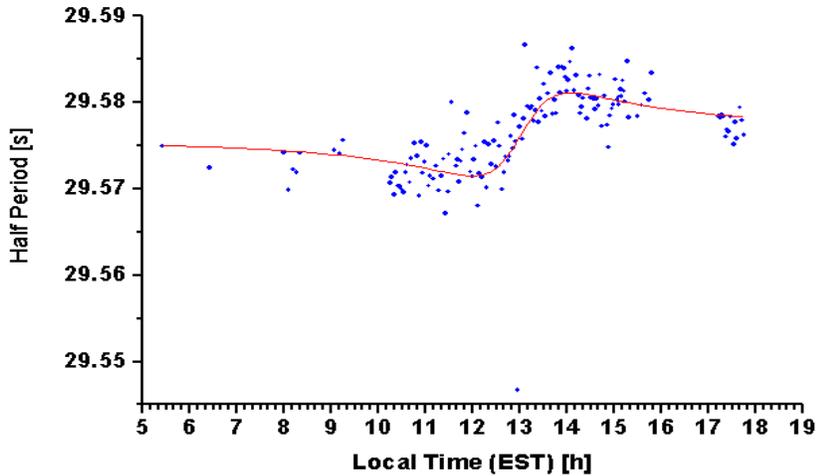

**Fig. 3.** A re-analysis of the Saxl and Allen data [Dui05]. All separate data points are used, Saxl and Allen combined 5 data points into one and used their standard deviation as value for the error bar. This has little influence on the fitted curve. EST (Eastern Standard Time) = UT – 5 h, also see [USNO]. First contact at 12:31 h, maximum 13:40 h, last contact 14:58 h. The totality at the observation point was ~96.5%.

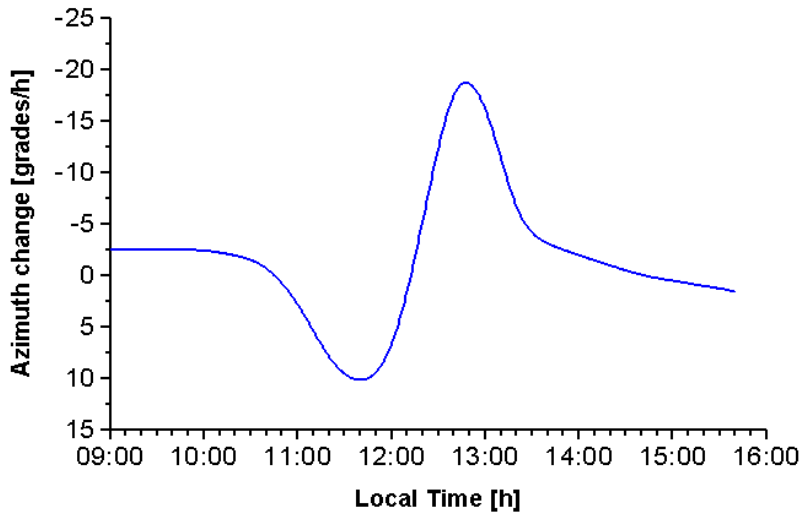

**Fig. 4.** Change in azimuth angle of the paraconical pendulum of Allais during the 1954 eclipse. Data fitted with a smooth curve and differentiated in order to yield the change per time. First contact at 11:21 h, maximum at 12:40 h, last contact 13:55 h (local time).

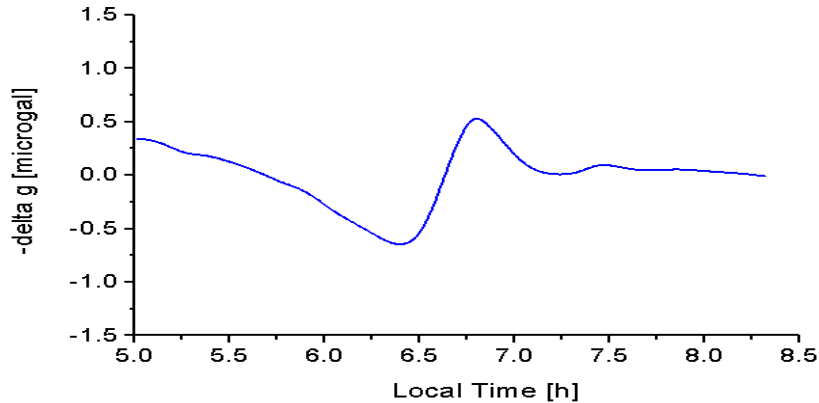

**Fig. 5.** Measurement by Mishra and Rao with a gravimeter during the 24 October 1995 eclipse in India [MisR97]. Deviations (inverted) from a linear trend curve of the de-tided gravity signal are shown. 1 μgal = $10^{-8}$ ms$^{-2}$. Maximum solar coverage between approx. 6:30 h and 7:00 h, according to the authors.

## 2. Conventional explanations of the observations

Explanations for the anomalous instrument readings during eclipses, based on conventional physics, can be subdevided into 6 kinds, which will be treated subsequently:
1. Seismic disturbances due to increased human activity before and just after an eclipse [UnnM01].
2. Gravitational effects of an increased density air mass spot due to cooling of the upper atmosphere [FlaY03].
3. Tilt due to temperature change of the soil and other atmospheric influences [YanW02, FlaY03].
4. Tilt due to atmospheric loading.
5. Influence of eclipse induced changes of the geomagnetic field.
6. Instrumental errors [RuyS03].

### 2.1. Increased seismic activity

Unnikrishnan et al. [UnnM01] developed a model based on the speculation that the gravity anomaly as observed by Wang et al. [WanY00] might be due to seismic disturbance caused by human activity. Preceding the first contact and after the last phase of the eclipse there will be increased traffic when people move into or out of the eclipse zone which causes the increase of seismic noise. In recent years, this hypothesis has been accepted a few times as a conventional explanation for the observations [*e.g.,* BerT02]. It is, however, highly unlikely to have been taken place during the measurements by Wang et al. Their measurements were performed in a geophysical observation station in a remote area in the North of China where the nearest road is several kilometers away and the experimenters were situated at 200 m distance [Yan04].

A survey of seismic data taken in Southern Belgium during the 1999 eclipse showed no sign at all of such an increased disturbance [Cam04]. Surely, if it was happening, it should be visible in such a densely populated area. Nevertheless, it is highly advisable that future measurements will be accompanied by the simultaneous measurement of seismic activity.

## 2.2. Gravitational effects of an increased density spot in the upper atmosphere due to cooling

Van Flandern and Yang modeled a traveling atmospheric excess mass spot due to cooling during the eclipse of higher atmospheric layers (above cloud level) [FlaY03].

Calculations show indeed that, when the density of the cooled spot of the upper atmosphere increases by 0.1% over the whole area of the shadow spot, a moving excess mass of the order of $10^{10} - 10^{11}$ kg is conceivable with its barycentre at an altitude of about 10 km. Taking the maximum value, this would yield a maximal vertical acceleration of

$$a_v = \frac{Gm_{air}}{(10 km)^2} \approx 7 \times 10^{-8} \text{ ms}^{-2},$$

indeed of the right magnitude. This model, however, presumes that air streams from the surrounding with speeds of the order of several 100 m/s. More realistic air speeds would imply that the "mass concentration" lags behind the shadow by at least 15 min (during which it will be already heated again, in thus way reducing the effect). The magnitude of this effect may therfore also be determined by the velocity of the shadow relative to the Earth surface. The magnitude of this varies from 0.5–1.0 km/s, depending on the eclipse.

The amount of barometric pressure increase at sea level of 0.6% which Van Flandern and Yang cite [Geo99] is a rather extreme value, which is confirmed nowhere. Van Ruymbeke et al. [RuyS03] report changes of the order of 0.5 hPa (~0.05%) in Vienna. Farges et al. [FarP03] report pressure changes due to the gravity waves[1] generated by the eclipse of the order of 10 Pa at sea level (=$10^{-5}$ effect). The observed pressure change at sea level should, actually, not be too important in the scenario described by Van Flandern and Yang. What is happening in this scenario is that the relation used for correction for atmospheric pressure change (the so-called admittance factor, *e.g.*, –0.3 µgal mbar$^{-1}$ [DucS99, CroH02]) breaks down.[2]

A problem is that little information is available about the behaviour of the atmosphere during eclipses at the altitudes used in these calculations. Virtually all publications about atmospheric effects either concern those near ground level or at high altitudes (ionosphere). Balloon measurements of pressure and temperature during solar eclipses at altitudes of the order of 10 km would therefore be very useful.

Also, in this model of Van Flandern and Yang, one would expect deviations of opposite sign around the observed deviations since there would necessarily be a depletion zone around the increased density spot. This is not visible in the data ([WanY00], Fig. 1) to a level ≤ 10% of the observed decreases in *g*.

Though this model still seems to be a reasonable candidate to explain the anomalous gravimeter behaviour, it certainly cannot account for the effects observed with torsion, paraconical and Foucault pendula, despite the title of the publication by Van Flandern and Yang. These devices have little sensitivity for changes in the vertical component of gravity. For changes of the magnitude of tidal forces (of the order of 100 µgal [Net76, DucS99]), Saxl and Allen determined the influence on the oscillation period of their torsion pendulum to be a factor of $10^5$ to small to account for their observations [SaxA71].

---

[1] A wave disturbance in the atmosphere in which buoyancy (or reduced gravity) acts as the restoring force on fluid parcels displaced from hydrostatic equilibrium. Not to confuse with gravitational waves as predicted by General Relativity.

[2] Gravimeter signals are often corrected for changes in atmospheric pressure [CroH02]. For the Chinese data no correction was applied. A change in pressure of approx. 1.3 hPa was observed for which the authors conclude that no correction was nescessary (effect << 1 µgal).

*2.3. Tilt due to temperature change of the soil and other atmospheric influences*

Cooling of the Earth crust due to the eclipse shadow has been mentioned a few times as being responsible for eclipse effects, *e.g.,* by Yang and Wang, who also studied it experimentally [YanW02]. They found a ground tilt < 10 µrad for a temperature drop of 2 K, whereas a tilt of the order of 100 µrad would explain the decrease in *g* of Wang et al. [WanY00] quantitatively.

On the qualitative side, it must be remarked that measurements of soil temperatures show that the drop in temperature due to an eclipse is only noticeable up to a dept of a few cm and that it lags the shadow by about 30 min [L-HY00, EatH97, FokW01] on sunny days. Furthermore, the magnitude of tilt is mainly determined by the gradient in temperature, which is usually not larger during an eclipse than during heating in the morning or cooling at the end of a day.

The measurements by Allais were performed in a basement, those by Saxl and Allen on a concrete floor which was attached to solid rock, which makes sensitivity due to temperature changes of these magnitudes unlikely.

Van Flandern and Yang [FlaY03] proposed that the setups of Allais and Saxl and Allen were sensitive to air currents due to pressure changes at the start of an eclipse. Literature about atmospheric effects during eclipses show that these effects are lagging behind the change of solar radiation and even that wind speed is reduced during eclipses [AndK75, FokW01, GirB01] and this conjecture is therefore not in occordance with the observations.

*2.4. Tilt due to atmospheric loading*

If a pressure increase occurs during the eclipse, as discussed in Section 2.2, this might also induce a tilt due to deformation of the Earth's crust. According to Rabbel and Zschau [RabZ85], maximum tilts of the order of 50 nrad are to be expected[3], to small to account for the observed effects (see also Section 3).

*2.5. Geomagnetic changes*

There exist a number of reports about eclipse induced changes of the geomagnetic field [Str01, BreL93, MalO00]. Athough this will be treated in more detail in the next version of this paper, preliminary analysis show that it cannot explain the anomalous effects qualitatively. The reported magnitudes are at most 10 nT and therefore of the same order as the diurnal changes of the geomagnetic field. Some articles even put geomagnetic changes at the Earth's surface into question [KorL01].

It will be useful, however, to calculate the effect of 10 nT changes on the various equipment mentioned in this paper (see, e.g., [CheC93], pp. 45–48 for calculating magnetic effects).

---

[3] Tilts due to tidal effects are of the order of 0.1 µrad.

*2.6. Instrumental errors*

A lot of reports about eclipse anomalies either lack a statistical analysis and/or records of sufficient length in time. A preliminary analysis of the Saxl and Allen data shows that their effect is a 3σ effect compared to normal daily changes in period [Dui05]. Also reports about measurements where no anomalous behaviour was found sometimes only show time windows which are barely larger than the eclipse event [Ruy03].

There are a multitude of causes which might seem to be able to explain disturbances of short duration, like one proposed by Van Ruymbeke et al.: *"An effect probably induced by some electrical occurrence due to the special situation of night conditions occurring during the day"* [RuyS03]. Although the data of, *e.g.,* Saxl and Allen [SaxA71] and Kuusela [Kuu92] also show some peaked deviations of many standard deviations magnitude, it must be concluded that causes like those proposed by Van Ruymbeke et al. cannot explain the general behaviour of the observations.

## 3. Observations of tilt

Interesting are observations of tilt during eclipses. Guérin and Gay [GueG99] already stressed the importance of these observations for resolving this enigma. Kuusela observed changes in the angle of his pendulum in the direction approximately perpendicular to the path of the shadow (*y*-direction) with a maximum of 2 μrad which *seem* to be related to the eclipse [Kuu92], see Fig. 6. No such effect was observed in the *x*-direction.

Latham describes measurements with a gyroscope during a solar eclipse in Western Australia in 1974 [Lat80]. He did not observe any effect in the gyroscope output, but did observe a tilt of the experimental platform in the E/W direction with a simultaneous operating tiltmeter of approx. 24 μrad. see Fig. 7. This is even an order of magnitude larger than the value observed by Kuusela. It was rather puzzling that the gyroscope did not show this rotation in its output. Latham also checked for changes in the value of *g* at the time of the eclipse in the records of the observatory at Mundaring, Western Australia, and found no indication of such an effect. It cannot be ruled out, however, that the observation was due to an instrument error.

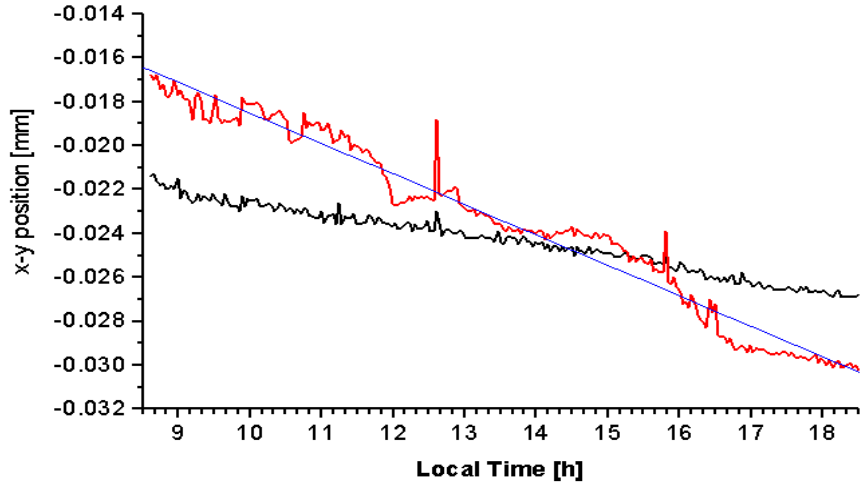

**Fig. 6.** Deviations of the midpoint of the torsion balance of Kuusela during the 1991 eclipse in Mexico [Kuu92]. Upper trace (red): *y*-direction, lower trace (black): *x*-direction. The *y*-direction is roughly in the direction of the path of the eclipse. The length of the pendulum wire was 1.00 m. First contact at 11:54 h, maximum from 13:21-13:27 h, last contact at 14:47 h local time.

An explanation could be that an extra horizontal force is sometimes generated during an eclipse. Free hanging "tilt meters" like those of Latham and Kuusela are sensitive to such a force while most gravimeters are not (at least, not for forces of this magnitude). This would also explain the lack of response of Latham's gyroscope, which was sufficiently sensitive to detect such a tilt [Lat80].

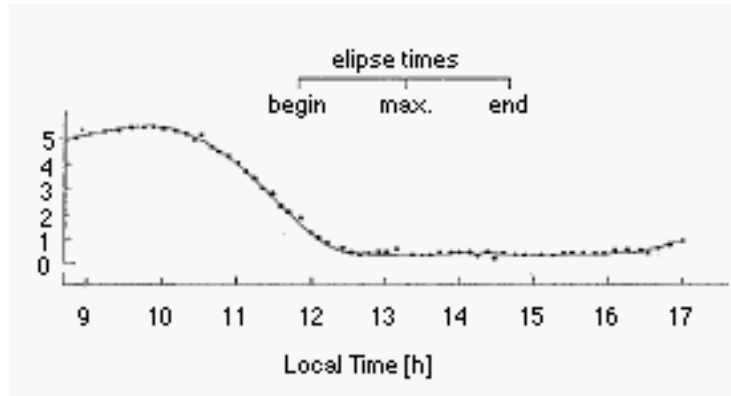

**Fig. 7.** Tilt as observed by R. Latham [Lat80] with a Talyval level during the June 20, 1974 solar eclipse in Western Australia. Vertical axis is the tilt in the East-West directions in sec of arc. Note that subtracting a linear trendline and inverting gives the same shape as Figs. 3, 4 and 5.

The observations of Kuusela show that the occurrence of tilt does not necessarily have an effect on the period of torsion pendulums. The Saxl and Allen pendulum had an other starting mechanism [SaxA80], however, so in principle it cannot be excluded that this device did have such a sensitivity to tilt. Saxl et al. have examined the sensitivity of their torsion pendulum to tilt a few years after the eclipse observations. Their conclusion was that horizontal movements of the pendulum disk of 0.1 mm would lead to a maximum change in period of 60 µs [SaxA80], while the observed changes during the eclipse are two orders of magnitude larger. If we take the magnitude found by Latham (5" ≈ 24 µrad), for the Saxl and Allen pendulum this would lead to a horizontal displacement of approx. 45 µm.

Guérin and Gay [GueG99] drew attention to the fact that the method of measurement of Allais with his paraconical pendulum[4] is in principle sensitive to tilt. Savrov, who did not observe anomalous behaviour with paraconical pendula[5], used a method less sensitive to this error[6] [Sav97]. The horizontal displacement for the Allais case will be at most 20 µm, whereas Allais could read the position on the outer dial to a precision not better than 200 µm. This seems to rule out tilt of this magnitude as an explanation for the observations of Allais.

**4. Conclusions and discussion**

Although, despite all proposed conventional explanations fail to explain the observations either qualitatively or quantitatively, it is still possible that the reported anomalies will turn out to be due to a combination of some of these effects and instrumental errors. And, of course, there may be yet unidentified conventional causes which play a role. The judgement of some of the experimental results is hampered by the lack of a statistical analysis and/or data of sufficient length. Nevertheless, there exist some strong data which cannot be easily explained away.

Therefore, further experiments during a few solar eclipses are justified. Measurements with a single device, however, seem to add little to the current situation. Combined measurements of several pendula and gravimeters together with monitoring of a variety of environmental parameters like seismic activity, ground tilt, atmospheric pressure, temperature, ground water level, etc. are necessary. Also, more study of the sensitivity of these instruments to the environmental factors described above, will be useful.

If an anomalous eclipse effect does exists, a clever combination of experiments and input from theoretical models will be necessary to reveal its existence. It seems clear that gravitational screening can be ruled out as theoretical explanation. Various studies have set upperlimits on this mechanism which rule out effects of the order of the eclipse anomalies [UnnM01, Eck90, Edw02, Fla96, SliC65].

There seem to be no studies published which estimate the magnitude of general-relativistic effects as possible cause for the observations. Comparison of, e.g., the magnitude of gravitomagnetic effects [RugT02] with the magnitude of the eclipse effects make it unlikely that the eclipse effects can be explained within the framework of General Relativity. Rather, it seems that one has to resort to gravitational theories which include preferred-location or preferred-frame effects (see, e.g., [Wil81], Section 6.3, for an account of these theories).

---

[4] Observing the change in azimuth for 14 min and releasing the pendulum at the last observed angle.

[5] From a recent article [Sav04], it can be inferred that he actually did observe an anomalous effect during the 1991 eclipse in Mexico, but ascribed it to atmospheric effects without justification.

[6] Savrov fitted the trajectory of the pendulum with an ellipse and measured the whole eclipse "with one release" of the pendulum.

**Acknowledgements**

The author wishes to thank Dr. Xinshe Yang (Univ. of Swansea, UK) for supplying additional data, Dr. Geoffrey Kolbe (Border Barrels, Scotland) for pointing him at the work of Robert Latham, Dr. Jay Burns (Florida Inst. of Technology) for correspondence, data and donating him the scientific archive of the late Erwin J. Saxl, Dr. Thierry Camelbeeck (ORB) for information about seismic activity in Belgium during the 1999 eclipse and Dr. Tom Kuusela for supplying his 1991 tilt data and additional information.

More information about this research will appear on: space-time.info